\documentclass[fleqn,usenatbib]{mnras}

\usepackage{amsmath}
\usepackage{mathptmx}
\usepackage{txfonts}
\usepackage[T1]{fontenc}
\usepackage{ae,aecompl}
\usepackage{graphicx}	
\usepackage{amssymb}	
\usepackage[normalem]{ulem}
\usepackage{longtable}
\usepackage{multirow}
\usepackage{lmodern} 
\usepackage{booktabs} 

\newcommand{\hmpc}{{\, h^{-1}\, {\rm Mpc}}}
\newcommand{\mpc}{{{\, \rm Mpc}}}

\def\apj{ApJ}
\def\apjs{ApJS}
\def\jcap{JCAP}
\def\mnras{MNRAS}
\def\aap{A\&A}

\def\nat{Nature}      
\def\apjs{ApJS}
\def\apjl{ApJ Letters}

\title[Longest filaments and the largest sheets in the cosmic web]{The maximum extent of the filaments and sheets in the cosmic web: an analysis of the SDSS DR17}

 \date{\today}

 \pubyear{2022}

 \author[Sarkar, Pandey  \& Sarkar]
         {Prakash Sarkar$^1$\thanks{E-mail:prakash.sarkar@gmail.com},
  Biswajit Pandey$^2$\thanks{E-mail:biswap@visva-bharati.ac.in} and
  Suman Sarkar$^3$\thanks{E-mail:suman2reach@gmail.com}\\
$^1$Department of Physics, Kashi Sahu College, Seraikella, Jharkhand - 833219, India\\
$^2$Department of Physics, Visva-Bharati University,
  Santiniketan, 731235, West Bengal, India\\$^3$Department of Physics, Indian Institute
  of Science Education and Research Tirupati, Tirupati - 517507,
  Andhra Pradesh, India}

\date{\today}

\begin{document}
\label{firstpage}
\pagerange{\pageref{firstpage}--\pageref{lastpage}}
\maketitle

\begin{abstract}
Filaments and sheets are striking visual patterns in cosmic web. The
maximum extent of these large-scale structures are difficult to
determine due to their structural variety and complexity. We construct
a volume-limited sample of galaxies in a cubic region from the SDSS,
divide it into smaller subcubes and shuffle them around. We quantify
the average filamentarity and planarity in the three-dimensional
galaxy distribution as a function of the density threshold and compare
them with those from the shuffled realizations of the original
data. The analysis is repeated for different shuffling lengths by
varying the size of the subcubes. The average filamentarity and
planarity in the shuffled data show a significant reduction when the
shuffling scales are smaller than the maximum size of the genuine
filaments and sheets. We observe a statistically significant reduction
in these statistical measures even at a shuffling scale of $\sim 130
\, \mpc$, indicating that the filaments and sheets in three dimensions
can extend up to this length scale. They may extend to somewhat larger
length scales that are missed by our analysis due to the limited size
of the SDSS data cube. We expect to determine these length scales by
applying this method to deeper and larger surveys in future.
\end{abstract}


\section{Introduction}
Quantifying the large-scale structures in the Universe and
understanding their origin is one of the central issues in
cosmology. The present-day universe exhibits structure over a wide
range of length scales. The observed structures like planets, stars,
galaxies, groups, clusters and superclusters show a clear hierarchy in
their order. In this hierarchy, the galaxies are the basic units of
large-scale structures. The observations from the modern redshift
surveys (SDSS \citealt{stout02}, 2dFGRS \citealt{colless01}) reveal
that the galaxies are distributed in an interconnected network of
filaments, sheets and clusters surrounded by gigantic voids, which is
often referred to as the ``cosmic web'' \citep{bond96}. The
observational evidence for such a complex network in the galaxy
distribution dates back to the late seventies and early eighties
\citep{gregory78,joeveer78,einasto80,zeldovich82,einasto84}. Understanding
the formation and evolution of the cosmic web has remained an active
area of research since then.

Filaments and sheets are prominent visual features of cosmic web. The
distribution of the galaxies are anisotropic on small scales due to
the geometry of these large-scale environments. The galaxy
distribution is also inhomogeneous due to wide variations in the
shape, size and density of these geometric patterns. However, the
inhomogeneity and anisotropy induced by these patterns are expected to
subside on large scales, provided the cosmological principle holds for
our Universe. The interconnected morphological components behave like
a nearly homogeneous network of galaxies on large-scales
\citep{sarkar19}. Such a transition can not occur below the length
scales of the largest sheets and filaments.

The assumption of statistical homogeneity and isotropy on large
scales, popularly known as the cosmological principle, is fundamental
to our understanding of the Universe. A large number of observations
support this assumption. Numerous studies
\citep{martinez94,borgani95,guzzo97,bharadwaj99,pan00,hogg05,yadav05,sarkar09a,scrim12,alonso15,pandey15,pandey16,sarkar16,avila19,goncalves21,pandey21}
indicate that our universe is statistically homogeneous on scales
beyond $70-150 \hmpc$. Studies with CMBR \citep{penzias65, smoot92,
  fixsen96} and a wide variety of tracers of the mass distribution
\citep{wu99,scharf00,blake02,gupta10,marinoni12,gibel12,yoon14,bengaly17,pandey17,sarkar18}
also indicate a transition to isotropy somewhere between $100-200
\hmpc$. However, the validity of the cosmological principle has been
challenged by several studies in recent years. Several studies report
the existence of large-scale structures that extend to several
hundreds of Mpc. Using the SDSS DR7, \citet{clowes13} report the
existence of a large quasar group spanning $\sim 500 \hmpc$ at $z \sim
1.3$. \citet{keenan13} identify a supervoid of diameter $\sim 600
\hmpc$ in the local universe. Recently, \cite{lopez22} report the
discovery of a giant structure of proper size 1 Gpc at $z \sim
0.8$. The existence of such giant structures in the universe suggests
inhomogeneity or anisotropy on surprisingly large scales which may be
in tension with the cosmological principle.  \citet{colin19} analyze
the JLA catalogue of Type Ia supernovae and find a bulk flow in the
local universe, which rejects the isotropy of the cosmic acceleration
at $3.9 \sigma$ statistical significance.  \citet{secrest21} analyze
1.36 million quasars from the WISE catalogue and find an anomalously
large dipole that contradicts the cosmological
principle. \citet{wiegand13} study the Minkowski functionals of the
luminous red galaxy (LRG) distribution in the SDSS and report greater
than $3 \sigma$ deviation from the $\Lambda$CDM model on scales of
$\sim 500 \hmpc$. Recently, \citet{appleby22} analyze the SDSS-III
BOSS data using the Minkowski functionals and find that the matter
density field is significantly anisotropic at low redshift. However,
they also point out that the north-south asymmetry in their analysis
may also arise due to some systematics in the data. All these
observations that point to apparent contradictions with the LCDM and
the cosmological principle are interesting in their own right and
deserve further attention and research.

Some known superclusters also seem to defy the assumption of
homogeneity and isotropy. The Sloan Great Wall, discovered by
\citet{gott05} in the SDSS, remains one of the strikingly large galaxy
systems that extend to length scales of more than $400 \, \mpc$. Our
home, the Milky Way itself, is part of the Laneakea supercluster
\citep{tully14} that extends to scales of $\sim 160 \,
\mpc$. \citet{lietzen16} discover a massive supercluster in the Baryon
Oscillation Spectroscopic Survey (BOSS) that consists of two walls
with diameters of $186 \hmpc$ and $173 \hmpc$ along with two other
superclusters of diameter $64 \hmpc$ and $91 \hmpc$. The Saraswati
supercluster \citep{bagchi17} is another wall-like structure at $z=3$
that spans at least $200 \, \mpc$. These superclusters are identified
in redshift space where the redshift space distortions may introduce
apparent structures that are not present in real space
\citep{praton97,melott98,shandarin09}. So the shape and size of these
superclusters may differ significantly in the redshift space and real
space. Further, the definitions of superclusters are often nebulous,
and their statistical significance as a single coherent structure
remains largely uncertain. The superclusters are generally dominated
by large filamentary and sheet-like structures
\citep{porter07,einasto14}. It is essential to test the statistical
significance of the sheet-like and filamentary patterns in the galaxy
distribution to assess the physical size of these large
superclusters. The maximum statistically significant size of the
sheets and filaments can also provide a lower bound on the scale of
homogeneity and isotropy.

\begin{figure*}
\resizebox{15cm}{6cm}{\rotatebox{0}{\includegraphics{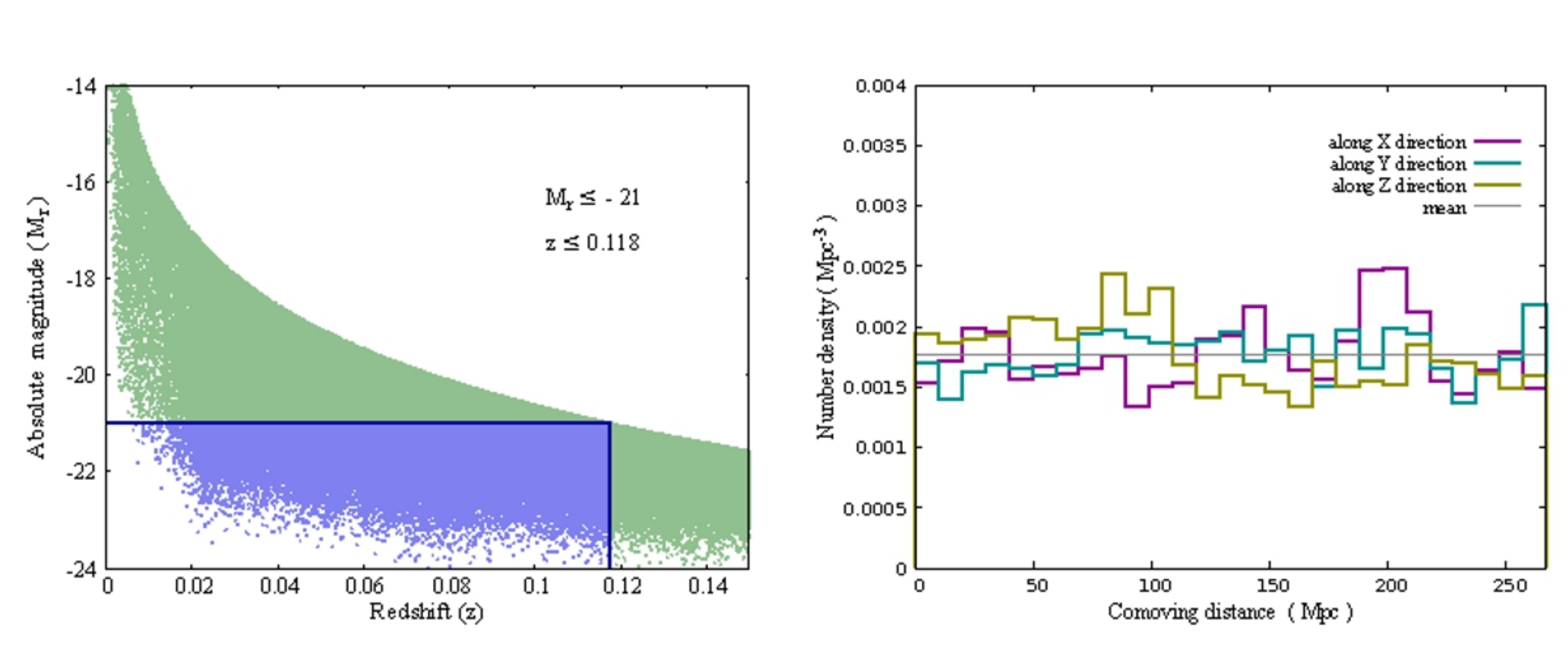}}}

\caption{The definition of the volume-limited sample in the
  redshift-absolute magnitude plane is shown in the left panel. The
  right panel shows the variation of comoving number density inside
  the SDSS data cube along the three co-ordinate axes. The densities
  are evaluated using slices of thickness of $10 \, \mpc$.}

\label{fig:sample}
\end{figure*}

Identifying the different morphological components of the cosmic web
and analyzing these clustering patterns is a challenging task. Various
statistical tools are developed for this purpose. Some of the measures
are the void probability function \citep{white79}, percolation
analysis \citep{shandarin83}, minimal spanning tree \citep{barrow85},
genus curve \citep{gott86}, Minkowski functionals \citep{mecke94},
multiscale morphology filter \citep{aragon07}, skeleton
\citep{novikov06}, spine \citep{aragon10} and the local dimension
\citep{sarkar09b}. The Minkowski functionals provide direct information
on the geometry and topology of the galaxy distribution. The ratio of
Minkowski functionals are used to construct Shapefinders
\citep{sahni98} that can describe the morphology of the large scale
structures.

The filaments and the sheets are anisotropic structures with a great
variety of shapes and sizes. It is generally difficult to define these
structures and determine their size due to their structural variety
and complexity. Some earlier works quantify the statistically
significant length scale of filaments in the SDSS Main Galaxy sample
\citep{bharadwaj04, pandey05, pandey10} and the Luminous Red Galaxy
(LRG) sample \citep{pandey11}. The results show that the filaments are
statistically significant up to $80 \hmpc$ in the SDSS Main Galaxy
distribution and $110 \hmpc$ in the LRG distribution. These studies
are based on the analysis of the two-dimensional projections of the
three-dimensional galaxy distribution. The projection effects in such
samples are expected to introduce spurious patterns. Some of the
larger filaments in such sample may arise due to the projection of
multiple filaments and sheets. Further, the sheet-like structures can
not be studied reliably in such projected galaxy samples. Keeping
these in mind, we plan to analyze the three-dimensional galaxy
distribution from the SDSS to study the size of the longest filaments
and the largest sheets in the cosmic web. It would determine the
physical scale up to which the large-scale structures are coherent in
a statistically significant way.

\citet{sheth03} develop the SURFGEN code to quantify the geometry and
topology of the large-scale structures in the cosmic web. The method
is used to study the morphology of the large-scale structures in
simulations \citep{shandarin04} and mock galaxy catalogues
\citep{sheth04}. The SURFGEN employs the `{\it{Marching Cube}}'
\citep{lorenson87} algorithm for triangulation and surface
modelling. An advanced version of SURFGEN is developed by
\citet{bag18} and named SURFGEN2. SURFGEN2 triangulates the surfaces
by using the `{\it{Marching Cube} 33}' \citep{chernyaev95}
algorithm. SURFGEN2 determines the Minkowski functionals and the
Shapefinders for each identified structure. We plan to use SURFGEN2 to
measure the filamentarity and planarity of the large-scale structures
in three-dimension and assess their significance using a statistical
technique `{\it Shuffle}' \citep{bhavsar88}.

A brief outline of the paper follows. In section 2, we describe the
data followed by the method of analysis in section 3. We present the
results in section 4 and conclusions in section 5.

Throughout the paper, we use the $\Lambda$CDM cosmological model with
$\Omega_{m0}=0.315$, $\Omega_{\Lambda0}=0.685$ and $h=0.674$
\citep{planck18} for conversion of redshift to comoving distance.

\section{DATA} 

\subsection{SDSS DR17 data}
We use data from the seventeenth data release (DR17) of the Sloan
Digital Sky Survey (SDSS) for the present analysis. The SDSS is the
largest redshift survey to date. It is a multi-band imaging and
spectroscopic redshift survey that uses a 2.5 m telescope at Apache
Point Observatory in New Mexico. A technical description of the SDSS
telescope is provided in \citet{gunn06} and a technical summary of the
survey can be found in \citet{york00}. The SDSS photometric camera is
described in \citet{gunn98} and the selection criteria for the SDSS
Main Galaxy sample is discussed in \citet{strauss02}.

The SDSS, in its fourth phase, has mapped over 14 thousand square
degrees of the sky, targeting about 500 million unique and primary
sources, including over 200 million galaxies. For this study, we
extract data from the seventeenth data release (DR17)
\citep{abdurrouf22} of the SDSS. A structured query is run in the SDSS
CasJobs \footnote{https://skyserver.sdss.org/casjobs/} to extract the
required data.

We set the $scienceprimary$ flag to unity to ensure that only the
targets with the best spectrum are chosen. The $zWarning$ flag is set
to zero to select only the galaxies with a reliable redshift. A nearly
uniform region is chosen within the right ascension and declination
ranges $130^\circ \leq \alpha \leq 230^\circ$ and $0^\circ \leq \delta
\leq 62^\circ$ respectively. We prepare a magnitude limited sample
with the extinction corrected {\it r}-band apparent Petrosian
magnitude $<17.77$ and $k$-corrected {\it r}-band Petrosian absolute
magnitude limit $M_r \leq -21$ within the redshift range $ 0.01 \leq z
\leq 0.118$. The sample contains a total $113294$ galaxies. The
definition of our volume-limited sample is shown in the left panel of
\autoref{fig:sample}. The resulting sample has a radial extent of $
507.85 \,\mpc$ and it covers a total volume of $ 6.73 \times 10^{7} \mpc
^{3}$ with a mean number density of $1.68 \times 10^{-3} \mpc^{-3}$.

We finally carve out a cubic region from the volume-limited sample for
our analysis. We consider the largest cube that can be extracted from
the volume-limited sample. It provides us with a total $33648$
galaxies distributed within a cubic region of side $267 \mpc$. We show
the variation of the comoving number density in this cubic region
along three co-ordinates axes in the right panel of
\autoref{fig:sample}.

\begin{figure*}
\resizebox{15cm}{15cm}{\rotatebox{0}{\includegraphics{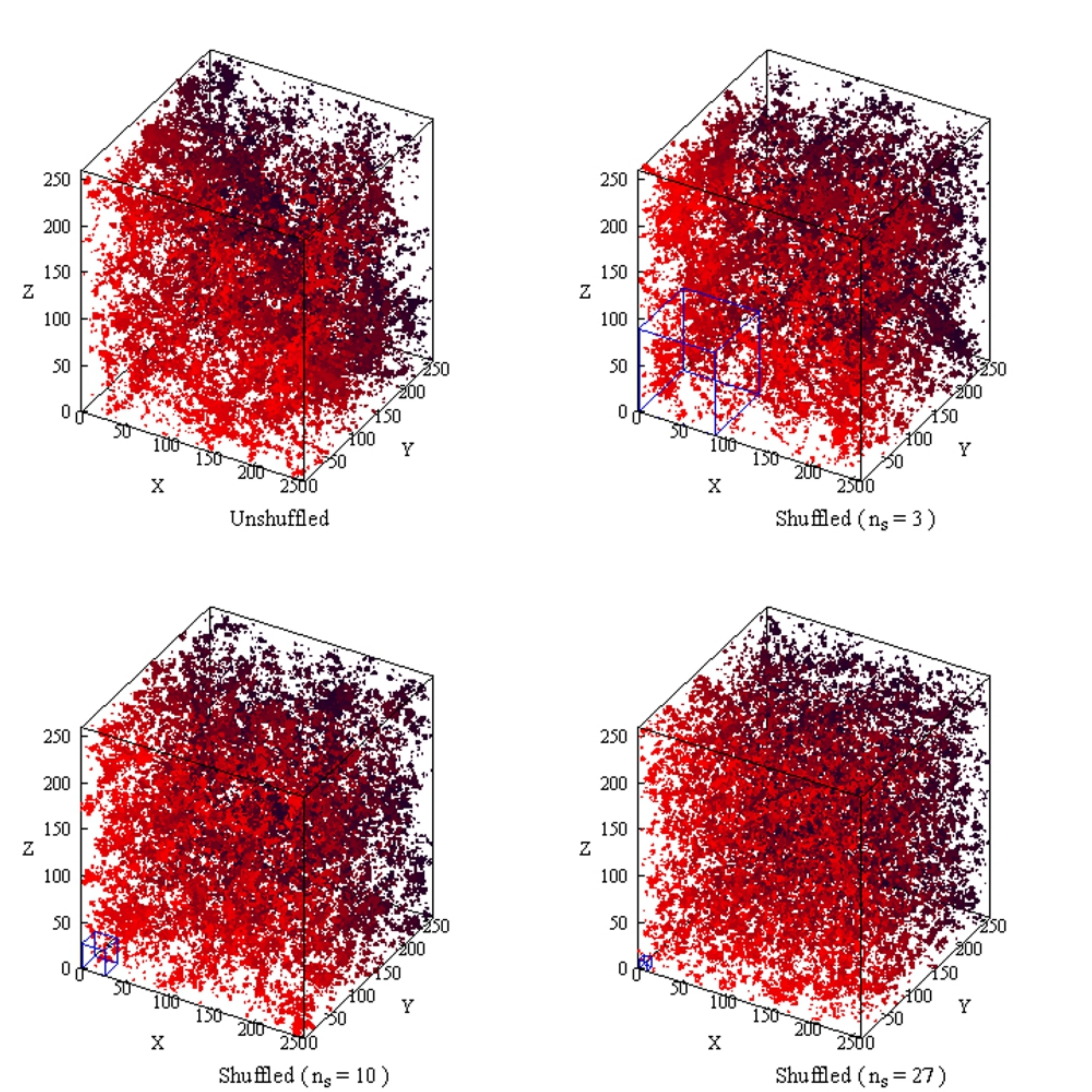}}}

\caption{The top left panel shows the distribution of galaxies in the
  SDSS data cube. The other panels show one realization of the
  shuffled datasets for different $n_s$ values as mentioned in each
  panel. A shuffling unit corresponding to each $n_s$ value is shown
  (in blue) at the corner of the respective shuffled data cube.}

\label{fig:shuffle}
\end{figure*}

\section{METHOD OF ANALYSIS}   

\subsection{Shuffle}

We subdivide the original data cube of side $267\,\mpc$ into
$N_c=N_s^3$ smaller sub-cubes of size $l_s \, \mpc$. The sub-cubes are
rotated around any of the three axes by a random angle which is
integral multiple of $90^{\circ}$. The axes of rotations are also
decided randomly. The spatial positions of the rotated subcubes are
then randomly interchanged. We repeat the random rotation followed by
swapping for a total $100 \times N_{c}$ times to generate a shuffled
version \citep{bhavsar88} of the original galaxy distribution within
the SDSS data cube. We label this procedure as `{\it shuffling}' and
term $l_s$ as the `{\it shuffling scale}'. The shuffling of the
original data is carried out for five different values of $n_s$. We
choose the following values of $n_s$: $2,\,3,\,5,\,10,\,27$ that
correspond to shuffling length of $133.5 \, \mpc,\,89 \, \mpc,\,53.4
\, \mpc,\,26.7 \, \mpc$ and $9.89 \, \mpc$ respectively. The galaxy
distributions in the original SDSS data cube and one shuffled
realization for three different shuffling scales are shown in
\autoref{fig:shuffle}.

\subsection{Minkowski functionals}

The Minkowski functionals (MFs) \citep{mecke94} can accurately
describe the three-dimensional morphology of a closed two-dimensional
surface. The four Minkowski functionals in three-dimensions are $-$
surface area ($S$), volume ($V$), integrated mean curvature ($C$), and
integrated Gaussian curvature or Euler Characteristics ($\chi$). The
first three MFs ($V, \,S$ and $C$) are geometric measures, while the
fourth ($\chi$) is a topologically invariant measure. The surface area
$S$ of the isodensity surface and the volume $V$ enclosed by it have
the usual meaning. The Integrated mean curvature $C$ of the isodensity
surface is defined as,
\begin{equation}
C = \tfrac{1}{2} \oint \big(\tfrac{1}{r_{1}}+\tfrac{1}{ r_{2}}\big) \, dS
\label{eq:imc}
\end{equation}
and the Euler characteristics $\chi$ of the isodensity surface is defined as,
\begin{equation}
\chi = \tfrac{1}{2 \, \pi} \oint \big(\tfrac{1}{r_{1}\, r_{2}}\big) \, dS
\label{eq:genus}
\end{equation}
where $r_{1}$ and $r_{2}$ are the principal radii of curvature at a
point on the surface.

We estimate these MFs for the isodensity surfaces defined at different
density thresholds. The distribution of the galaxies in the data cube
is first converted to a density field on a rectangular grid of size $1
\, \mpc$ using the Cloud-in-Cell (CIC) scheme. The resulting discrete
density field is then smoothed with a Gaussian filter of width $2 \,
\mpc$. The set of isodensity surfaces is extracted from the smoothed
density field at specific density thresholds ($\rho_{th}$). The
isodensity surfaces bound the structures with density above the
specific threshold. The individual structures are then identified with
a grid version of the friend-of-friend (FOF) algorithm
\citep{davis85}. We employ SURFGEN2 \citep{bag18, bag19} to define the
structures at each density threshold and calculate the associated
MFs. For each structure recognized by FOF, SURFGEN2 constructs a
closed triangulated surface using the Marching Cube 33 algorithm
\citep{chernyaev95}. We determine the MFs on the triangulated surface
using the following formula.

\begin{itemize}
\item $S = \sum_{i=1}^{N_T} S_i$, where $S_i$ is the area of the
  $i^{th}$ triangle and $N_T$ is the total number of triangles on the
  surface.\\

\item $V = \sum_{i=1}^{N_T} S_i(\hat{n_i}.\vec{P_i})/3$, where $\hat{n_i}$
  and $\vec{P_i}$ are the normal and centroid position vectors of
  $i^{th}$ triangle respectively.\\ 

\item $C=\sum_{i, j}\epsilon\,l_{ij} \, \phi_{ij}$, where $l_{ij}$ and
  $\phi_{ij}$ are the common edge and angle between the normals of
  adjacent triangles, respectively. The local concavity and convexity
  of the surface is encoded in $\epsilon$ with values of $\epsilon =
  -1$ and $\epsilon = 1$ respectively.\\

\item $\chi = N_T - N_E + N_V$ where $N_T, \,N_E$ and $N_V$ are
  respectively the total number of triangles, triangle-edges and
  triangle-vertices defining the isosurface.\\
\end{itemize}

\subsection{Shapefinders}
Using the MFs, \citet{sahni98} introduces the Shapefinders to quantify
the morphology of the large-scale structures in the galaxy
distribution. They are the thickness $\mathcal {T} =3V/S$, breadth
$\mathcal {B}=S/C$, and length $\mathcal {L} =C/4\pi$. All these
measures have a dimension of length. These dimensional shapefinders
and the Euler characteristics provide the typical size and topology of
the structures. A set of dimensionless shapefinders, namely the
filamentarity ($\mathcal{F}$) and planarity ($\mathcal{P}$), for each
structures, are defined as follows:\\
\begin{equation}
\mathcal{F} = \frac{\mathcal{L}-\mathcal{B}}{\mathcal{L}+\mathcal{B}},
\, \mathcal{P} =
\frac{\mathcal{B}-\mathcal{T}}{\mathcal{B}+\mathcal{T}}
\end{equation}\\
Here $\mathcal{F}=1,\, \mathcal{P}=0$ represents an ideal filament,
whereas $\mathcal{F}=0,\, \mathcal{P}=1$ is for an ideal sheet. For the
purpose of studying the general morphology of the galaxy distribution,
we define average filamentarity and average planarity as the
volume-weighted sum of the filamentarity and planarity of individual
structures. They are defined as follows:\\
\begin{equation}
F_a = \frac{\sum_i V_i \mathcal{F}_i}{\sum_i V_i}, \, \, 
P_a = \frac{\sum_i V_i \mathcal{P}_i}{\sum_i V_i}
\end{equation}\\
where $V_i$ is the volume of $i^{th}$ structure identified by FOF for
a given density threshold ($\rho_{th}$). The summation is across all
such structures recognized by FOF for that $\rho_{th}$. Consequently,
the structure with the highest volume contributes most to the $F_a$
and $P_a$. The largest structure grows in volume as the density
threshold is lowered. The growth of the largest structure can be well
captured by the Largest Cluster Statistic (LCS). The LCS is defined as
the fraction of volume occupied by the largest structure. We calculate
the LCS as,\\
\begin{equation}
LCS= \frac{V_{LS}}{\sum_i V_i}, 
\end{equation}\\
where $V_{LS}$ is the volume of the largest structure.

We evaluate $F_a$, $P_a$ and the LCS for the original unshuffled SDSS
data cube and five sets of shuffled data cubes at various density
thresholds ($\delta \in [-0.8, 12]$). Each shuffled set consists of
ten realizations that are used to estimate the mean and $1 \sigma$
errorbars for our measurements.

\begin{figure*}
\resizebox{18cm}{12.86cm}{\rotatebox{0}{\includegraphics{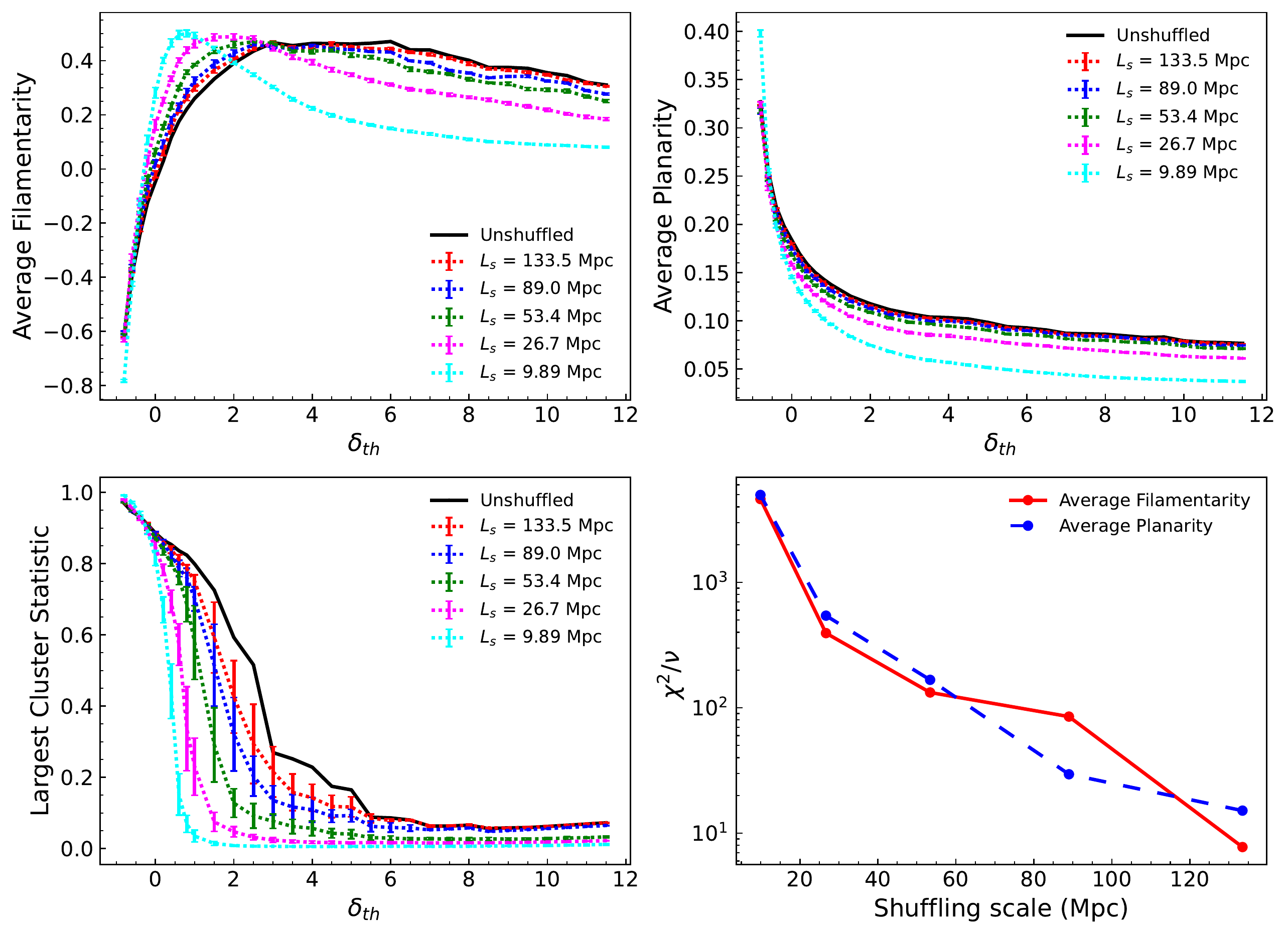}}}
\caption{The top left, top right and bottom left panels of this figure
  respectively show the average filamentarity, average planarity and
  largest cluster statistic in the original unshuffled data and its
  shuffled versions as a function of the density threshold. The
  largest cluster statistic is the fractional volume occupied by the
  largest cluster. The shuffling scale associated with each shuffled
  version are indicated in each panel. The $1 \sigma$ error bars at
  each data point are obtained from 10 different realizations for each
  shuffling scale. The errorbars for the average filamentarity and
  planarity are noticeably small. We quantify the difference between
  the original and the shuffled data as a function of the shuffling
  scale in the bottom right panel.}

\label{fig:allfigure}
\end{figure*}

We quantitatively compare the average filamentarity and average
planarity in the actual data to that with the shuffled datasets in the
bottom right panel of \autoref{fig:allfigure}. We calculate the means
$\bar{F_a}[{\it{Shuffled}}]$, $\bar{P_a}[{\it{Shuffled}}]$ and the
variances $(\Delta F_a[{\it{Shuffled}}])^2$, $(\Delta
P_a[{\it{Shuffled}}])^2$ for the average filamentarity and average
planarity at each $\delta_{th}$ using 10 realizations for each
shuffling scale. We quantify the difference between the average
filamentarity of the actual data and the shuffled data using the
$\chi^2$ per degree of freedom as,
\begin{equation}
\frac{\chi^2}{\nu}= \frac{1}{N_p-1}\sum_{i=1}^{N_p}\frac{(F_a[{\it{Actual}}]-\bar{F_a}[{\it{Shuffled}}])_i^2}{(\Delta
F_a[{\it{Shuffled}}])_i^2}
\end{equation}\\
where $N_p$ is the total number of density threshold used in the
analysis. We also estimate the $\chi^2$ per degree of freedom for the
average planarity as a function of shuffling scale in a similar
manner.

\begin{figure*}
\resizebox{7.5cm}{6cm}{\rotatebox{0}{\includegraphics{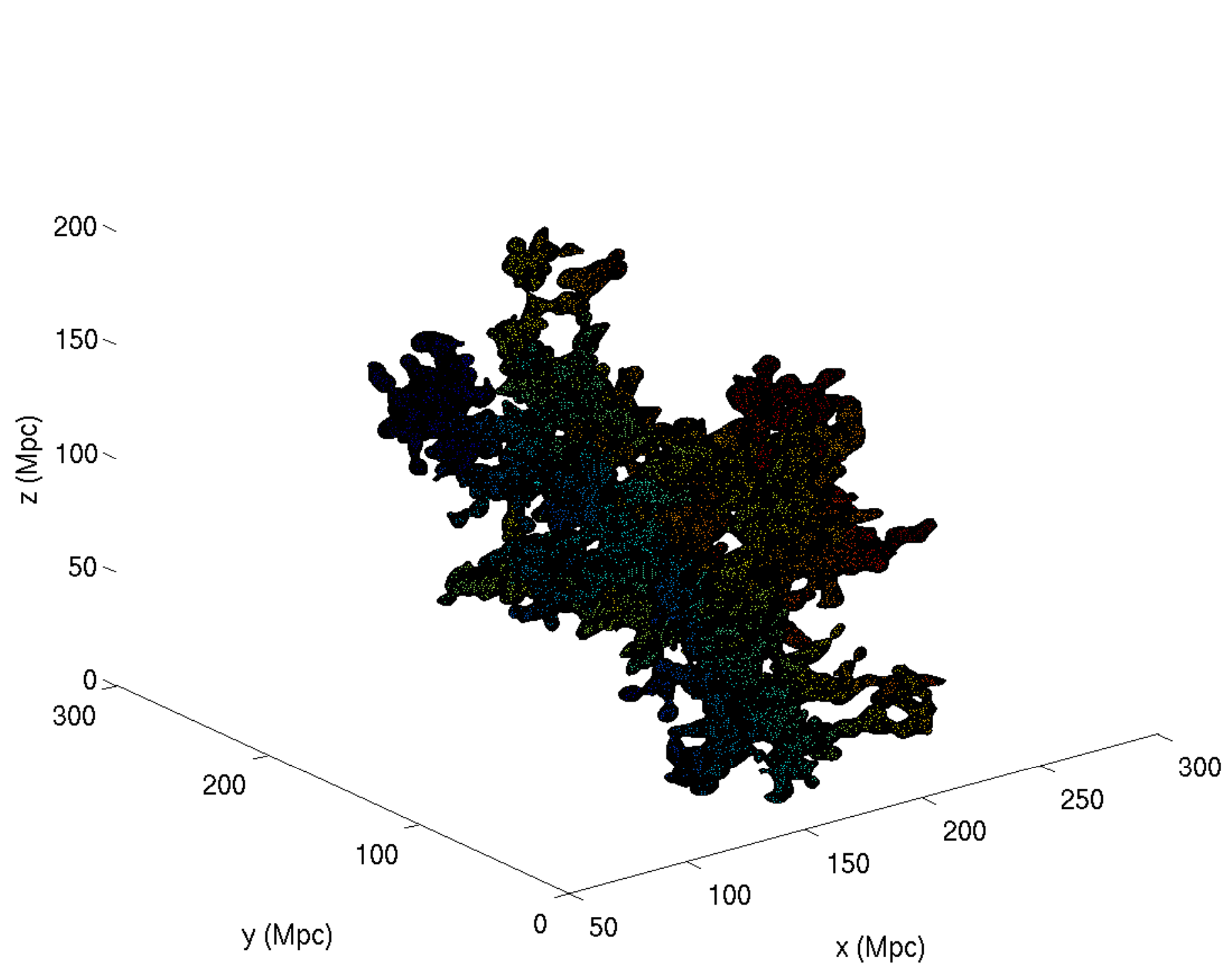}}} 
\resizebox{7.5cm}{6cm}{\rotatebox{0}{\includegraphics{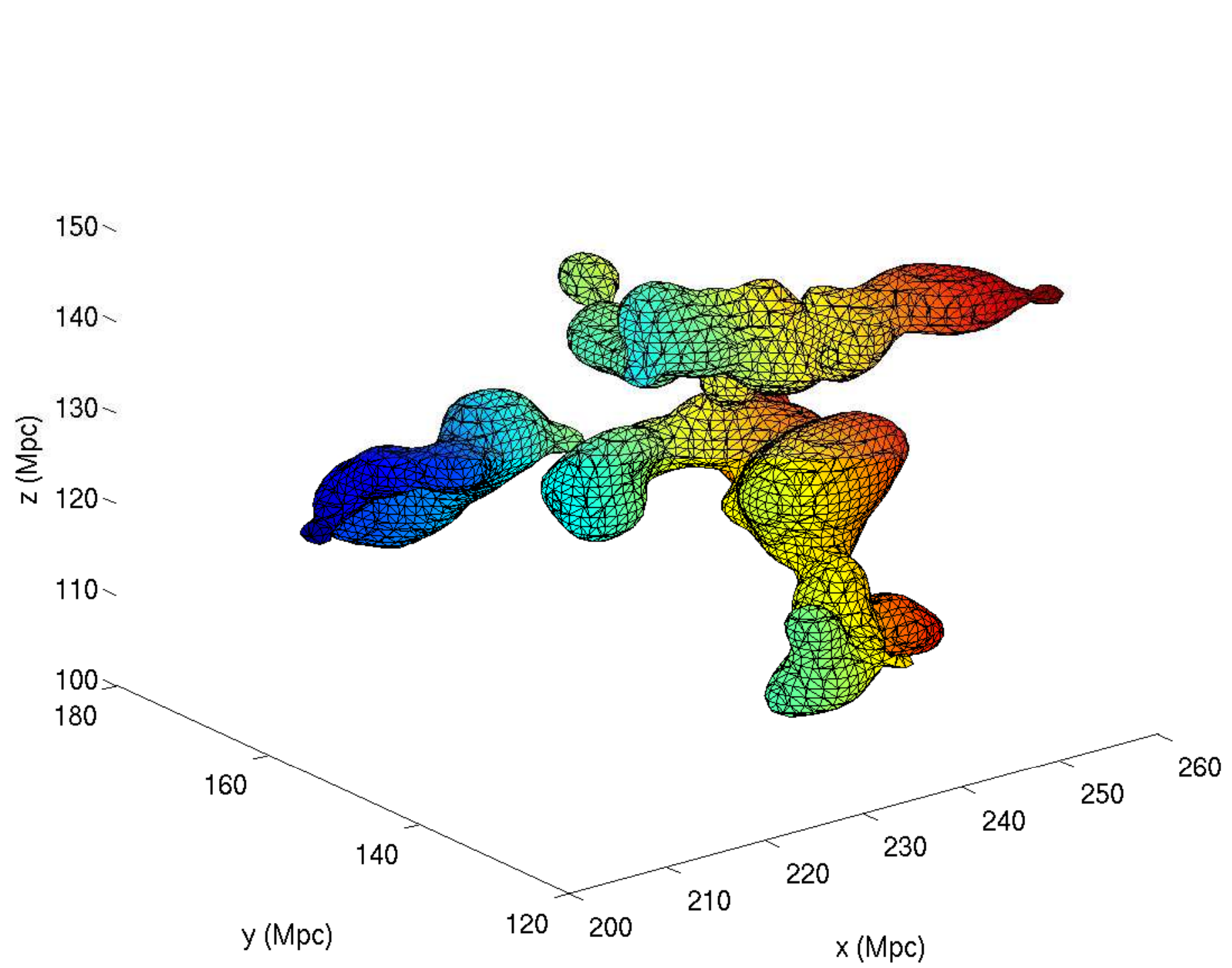}}}

\caption{The left panel shows the largest structure identified from
  the smooth density field of the SDSS galaxy distribution at a
  density threshold of $\delta_{th}=3$ before the onset of
  percolation. The right panel shows the largest structure in one of
  the shuffled realizations with a shuffling scale of
  $9.89\,\mpc$. The largest structure in the shuffled data is also
  identified at the same density threshold. It may be noted that the
  two panels cover different ranges of length scales. The relatively
  smaller size of the largest structure in the shuffled data at the
  same density threshold is evident in this figure. The figure clearly
  shows that the largest structure in the original data poses a higher
  degree of filamentarity and connectivity compared to the same in the
  shuffled data.}

\label{fig:largestc}
\end{figure*}

\section{RESULTS}

We analyze the SDSS galaxy distribution within the cubic region and
its different shuffled versions using SURFGEN2. The results of our
analysis are shown in different panels of \autoref{fig:allfigure}.

We show the average filamentarity ($F_a$) in the original unshuffled
galaxy distribution as a function of the threshold density contrast
($\delta_{th}$) in the top left panel of \autoref{fig:allfigure}. The
average filamentarity in the different shuffled versions of the
original galaxy distribution are also shown together in the same
panel. The $1 \sigma$ errorbars shown at each data point for the
shuffled data are obtained from 10 realizations for each shuffling
scale. We find that for both the unshuffled and the shuffled data, the
average filamentarity in the galaxy distribution slowly increases with
the decreasing density threshold and eventually reaches a maximum.
It then decays rapidly for a further decrease in the density
threshold. For the original unshuffled data, the average filmentarity
reaches a maximum at $\delta_{th} \sim 3$. The average filamentarity
drops to subzero values at $\delta_{th} < 0.1$.

Lowering the density threshold interconnects the individual
structures, producing even larger structures. The filamentarity of the
galaxy distribution keeps increasing until the percolation threshold
is reached. At percolation, the individual structures rapidly merge
into a connected structure spanning the entire volume. The density
threshold corresponding to this rapid transition is termed the
percolation threshold. The galaxy distribution assumes a sponge-like
topology after the onset of percolation, where the large-scale
structures interconnect into a giant network surrounded by immense
voids. After the percolation transition, the network becomes thicker,
and the voids turn rounder when the density threshold is lowered
further. The large near spherical voids surrounded by the filaments
introduce considerable negative curvature in the system. This negative
curvature is responsible for the sizeable negative filamentarity at a
very low-density threshold.

The average filamentarity in the shuffled data is significantly lower
than that in the unshuffled data at $\delta_{th} > 3$. It indicates
that the shuffled galaxy distributions are less fiamentary than the
original unshuffled data. The shuffled data with the smallest
shuffling scale shows the largest drop in the average
filamentarity. The average filamentarity in the shuffled data at each
density threshold beyond $\delta_{th}=3$ increases with the increasing
shuffling scale which clearly shows that the shuffling procedure
destroys many filaments in the galaxy distribution. The shuffling
destroys the filaments when the shuffling scale is smaller than the
size of the filaments in the data. At $\delta_{th} < 3$, we see a
reversal in the trend. The shuffled data have a higher average
filamentarity than the actual data simply because of the shift of the
percolation threshold in the shuffled data to lower density. It may be
noted that the average filamentarity peaks at a lower density
threshold for the shuffled datasets.

We show the average planarity ($P_a$) in the original unshuffled
galaxy distribution and its shuffled versions as a function of the
density threshold in the top right panel of
\autoref{fig:allfigure}. We find that the average planarity steadily
increases for both the unshuffled and shuffled data as we lower the
density threshold. After the percolation transition, the filaments
only grow thicker with decreasing density threshold. Consequently, the
large-scale structures assume an increasingly planar morphology at a
lower density threshold.

While comparing the average planarity of the shuffled data with that
from the original unshuffled data, we note that the average planarity
decreases at each density threshold with increasing shuffling
scale. The shuffled data with the smallest shuffling scale show the
most significant drop in the average planarity. These clearly show
that shuffling the data destroys the sheet-like structures in the
galaxy distribution. The differences between the average planarity of
the original and the shuffled datasets are more visible if we use a
logarithmic scale for the average planarity. Here we have plotted the
average planarity in a linear scale.

In other words, the difference between the average filamentarity or
the average planarity in the unshuffled and the shuffled data
decreases with the increasing shuffling scale. It can be understood as
follows. The original data may have coherent filaments and sheets up
to specific length scales. Shuffling the data would destroy all the
filaments and sheets spanning beyond the shuffling scale. However, the
filaments and sheets smaller than the shuffling length scale would
survive and remain nearly intact. If the most extended filaments or
the most extensive sheets are shorter than the shuffling length then
the average filamentarity or the average planarity in the shuffled
data would not show any significant changes from that in the original
data. Shuffling the data on a larger scale allows more filaments and
sheets to survive the shuffling process. At a given shuffling scale,
the average filamentarity or the average planarity in the original
data would be larger than in the shuffled data only if the original
data have more filaments or sheets spanning beyond the shuffling
length, than that expected from chance alignments. When the average
filamentarity or the average planarity in the original data are within
the $1 \sigma$ errorbars of that for the shuffled data then one can
infer that the maximum size of the filaments or sheets must be smaller
than the associated shuffling scale. This would only occur when the
longest filaments or the largest sheets are shorter than the
associated shuffling scale. Larger filaments or sheets may still exist
in the galaxy distribution but they are the product of pure chance
alignments.

\begin{figure*}
\resizebox{7.5cm}{6cm}{\rotatebox{0}{\includegraphics{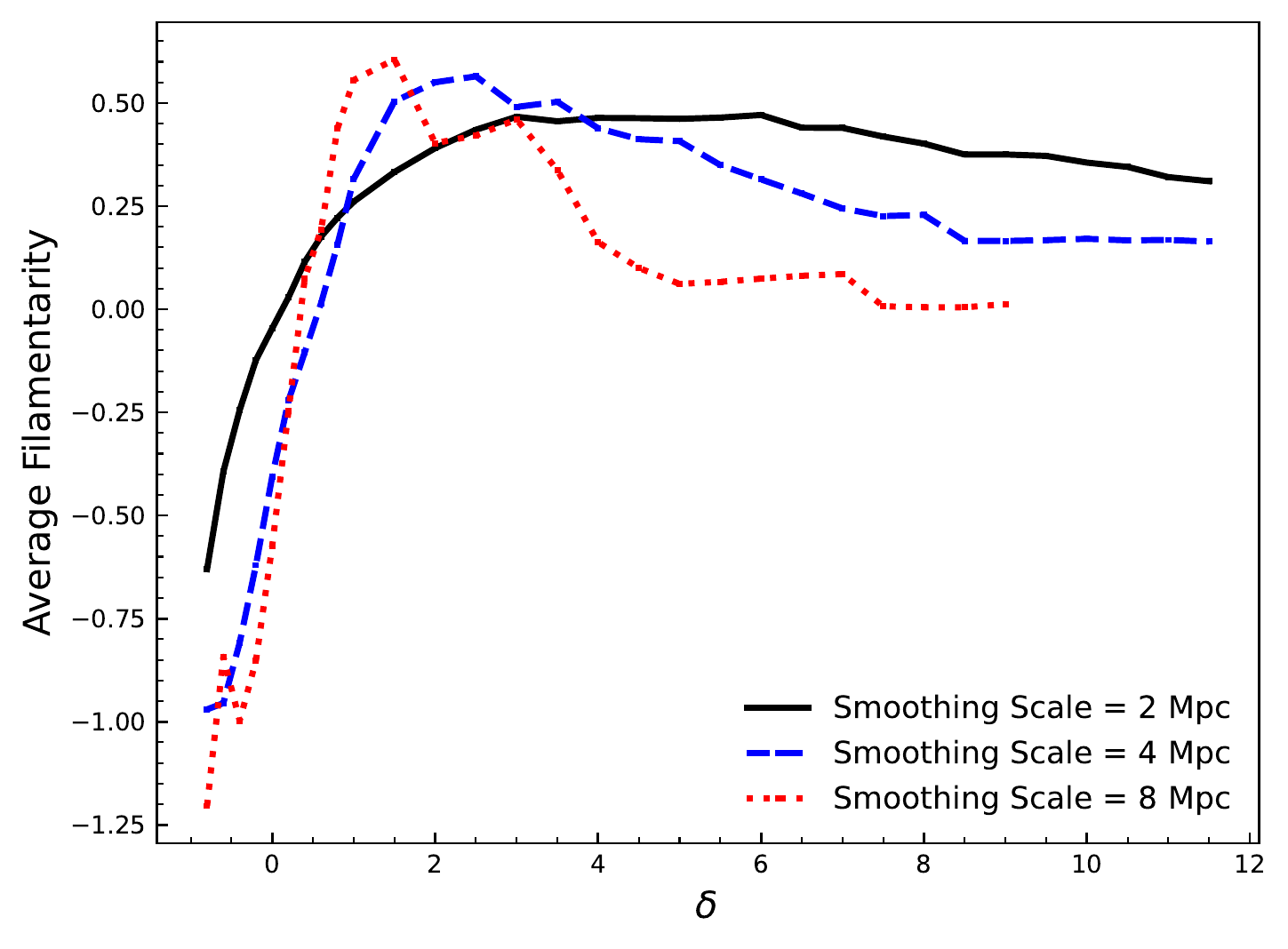}}} 
\resizebox{7.5cm}{6cm}{\rotatebox{0}{\includegraphics{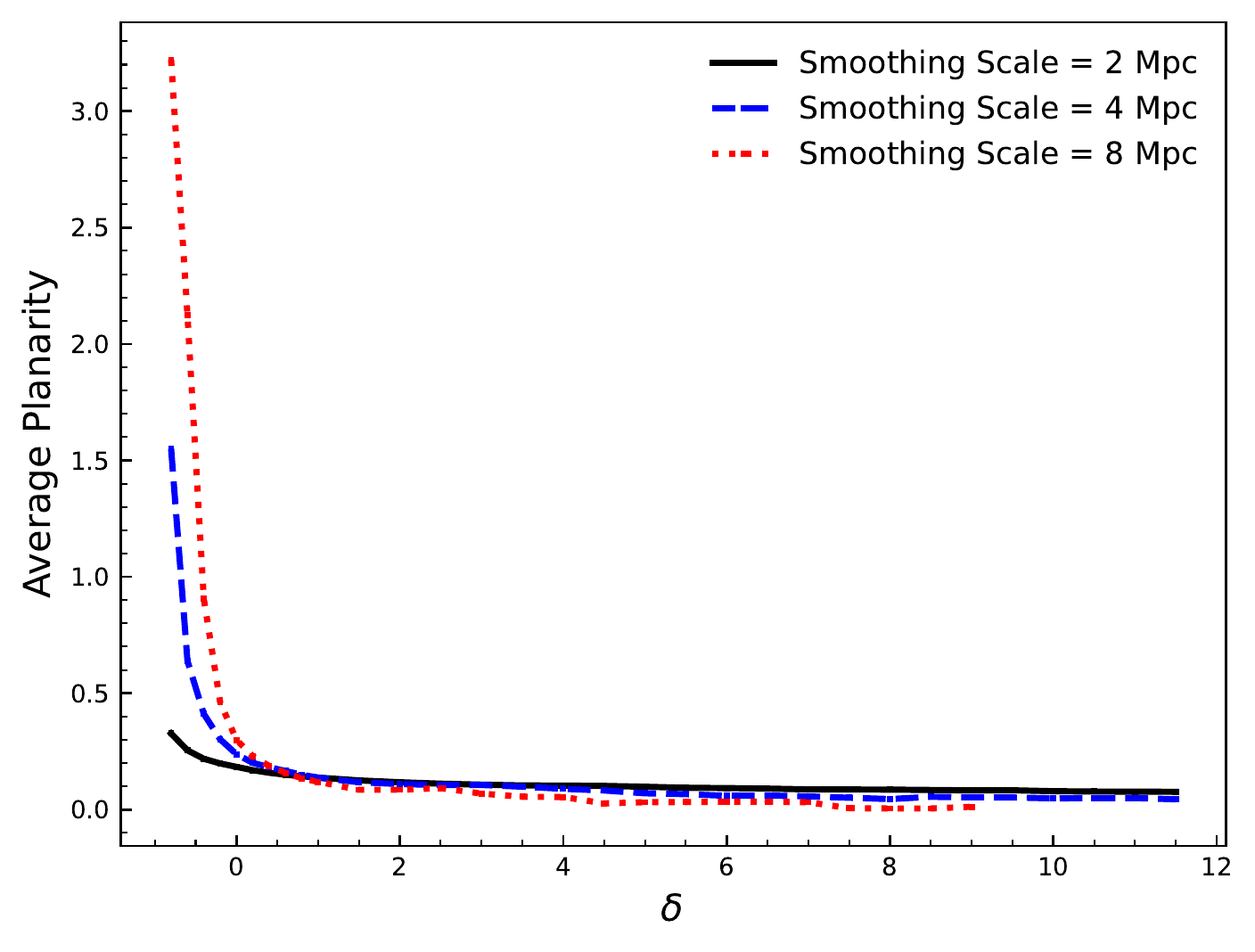}}}
\resizebox{7.5cm}{6cm}{\rotatebox{0}{\includegraphics{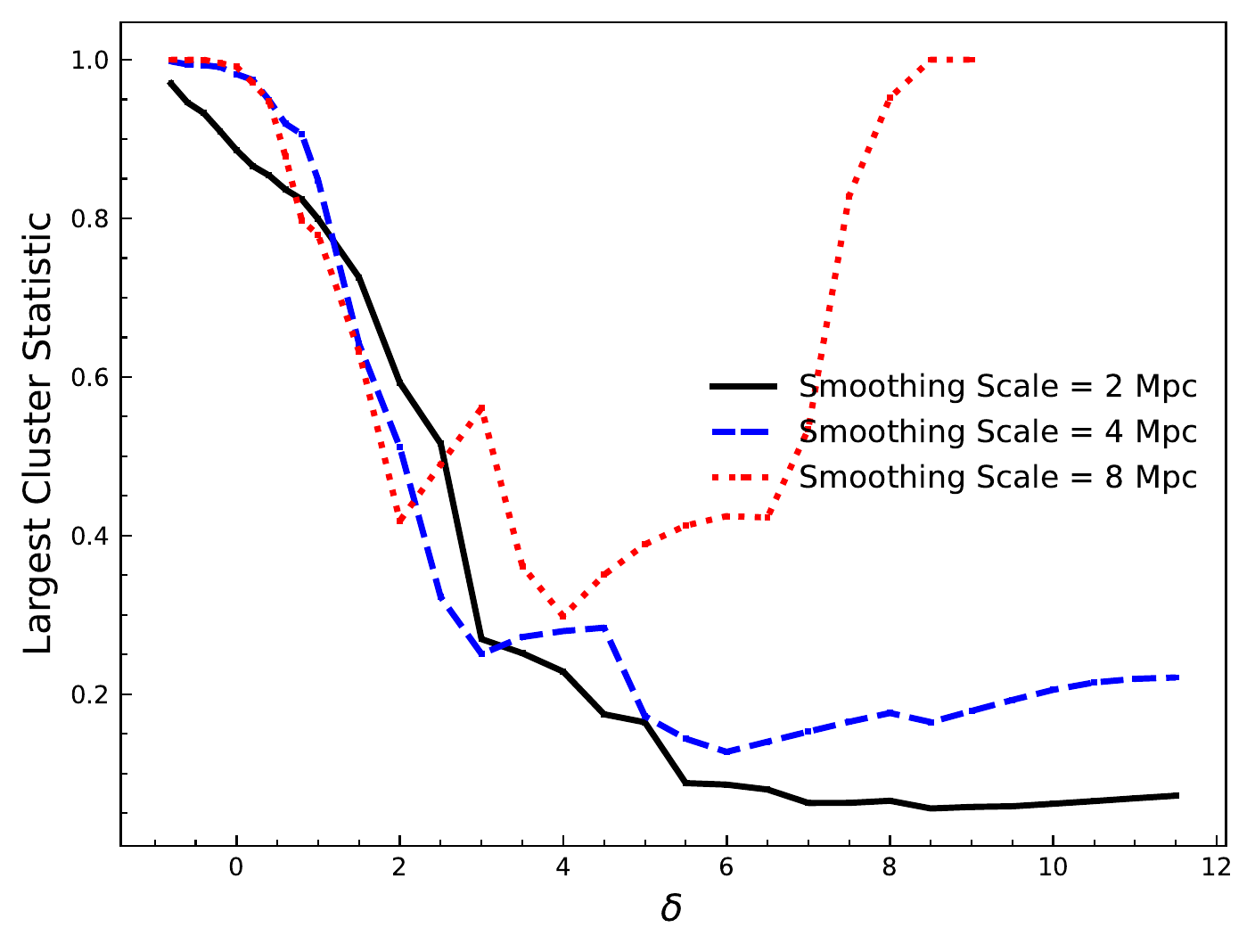}}}
\caption{The three panels of this figure show the average
  filamentarity, average planarity and the largest cluster statistic
  in the original SDSS data as a function of density threshold for
  three different smoothing lengths.  }
\label{fig:smoothing}
\end{figure*}

The bottom right panel of \autoref{fig:allfigure} shows the chi-square
per degree of freedom for the average filamentarity and the average
planarity, as a function of the shuffling scale. The chi-square per
degree of freedom is very large ($\sim 4.5 \times 10^3$) at the
smallest shuffling scale ($9.89 \, \mpc$) for both the average
filamentarity and average planarity. It suggests that the filaments
and sheets are highly significant at this length scale. In both cases,
the $\chi^2/\nu$ gradually decreases with the increasing shuffling
scale. We note that it decreases to a value of $\sim 8$ for the
average filamentarity and $\sim 15$ for the average planarity at the
shuffling scale of $133.5 \, \mpc$. So the average filamentarity and
the average planarity in the shuffled data differ from the actual data
in a statistically significant way, even at the largest shuffling
scale of $133.5 \, \mpc$. It implies that the filaments and sheets in
the galaxy distribution are statistically significant up to the
largest length scale probed in this analysis. Ideally, a
$\frac{\chi^2}{\nu} \sim 1$ would indicate the filaments or sheets
span up to the same length scales in the original and shuffled data
and no structures are destroyed by the shuffling
procedure. Unfortunately, we can not probe the length scales beyond
$133.5 \, \mpc$ with the existing data. Shuffling the data on somewhat
larger scales ($150-200 \, \mpc$) would most likely eliminate all the
differences observed between the actual and the shuffled data. A
larger value of the chi-square per degree of freedom for the average
planarity compared to the average filamentarity at the largest
shuffling scale indicates that the largest sheets may extend to
somewhat larger length scales than the longest filaments.

The average filamentarity and planarity are mostly contributed by the
structures with larger volumes. Naturally, the largest structure at
any given density threshold contributes more to these statistics. We
show the Largest Cluster Statistic (LCS) as a function of the density
threshold for the original and shuffled data in the bottom left panel
of \autoref{fig:allfigure}. The LCS quantifies the fraction of volume
occupied by the largest structure and shows the growth of the largest
structure with decreasing density threshold. A faster growth of the
LCS indicates greater connectivity of the large-scale structures in
the galaxy distribution. The bottom left panel of
\autoref{fig:allfigure} shows that the LCS grows slowly in the
original SDSS data as the density threshold is lowered. The LCS in the
SDSS data shows a sudden jump from $0.25$ to $0.75$ when the density
threshold ($\delta_{th}$) is decreased from 3 to 2. The sudden jump in
the LCS corresponds to the percolation transition. The LCS in all the
shuffled data sets are clearly smaller than the actual data for
$\delta_{th}>0$. The jump in the LCS occurs at a smaller density
threshold for the shuffled data, indicating a shift in their
percolation threshold to the lower densities. The LCS in the shuffled
data decreases at each density threshold with the decreasing shuffling
scale. We note that for the smallest shuffling scale, the LCS stays
$\sim 0.02-0.05$ up to $\delta_{th} \sim 1$ and then suddenly jumps to
$0.8$ at $\delta_{th} \sim 0.5$. It clearly shows that shuffling the
data decreases the connectivity of the galaxy distribution by
destroying large-scale patterns like sheets and filaments. The results
shown in this panel also assert that shuffling the data on a scale of
$133.5 \, \mpc$ diminish the connectivity in the galaxy distribution
in a statistically significant way. It is consistent with the results
obtained from the analysis of the average filamentarity and planarity.

We show the largest structure before the onset of percolation in the
original data at $\delta_{th}=3$ in the left panel of
\autoref{fig:largestc}. The largest structure identified from one of
the shuffled realizations with shuffling scale $9.89 \, \mpc$ is shown
for comparison in the right panel of \autoref{fig:largestc}. The
largest structure in the shuffled data shown in this panel is also
identified at $\delta_{th}=3$. It may be noted that the two panels
cover different ranges of length scales, and the largest structure in
the shuffled data is noticeably smaller compared to the original data
at the same density threshold. The largest structure in the shuffled
data is a simply connected object, whereas the one in the original
data is a multiply connected complex object with a higher degree of
filamentarity and connectivity.

The results for the SDSS data in \autoref{fig:allfigure} hint a
critical density threshold at $\delta \sim 3$ below which most of the
individual structures interconnect to become one single network-like
structure spanning the entire volume. This corresponds to the
percolating density threshold. We test if this threshold depends on
the smoothing length chosen for the analysis. We compare the average
filamenarity, average planarity and the largest cluster statistic for
3 different smoothing lengths in \autoref{fig:smoothing}. We find that
the percolating density threshold is sensitive to the smoothing
length. We analyze the SDSS data with smoothing lengths of 2 Mpc, 4
Mpc and 8 Mpc and note that the percolating density thresholds
decreases with the increasing smoothing length. The results for
different smoothing lengths are qualitatively similar. We notice that
for the smoothing length of 8 Mpc, LCS shows a strikingly different
behaviour where it initially decreases and then increases with the
decreasing density threshold. A higher smoothing would lead to lesser
number of high density regions. So there will be very few structures
at higher density thresholds. The LCS would be very high ($\sim 1$)
when the largest among these few structures is significantly larger
than the rest. The LCS initially decreases with the increase in the
available number of structures as the density threshold is
lowered. However, these structures would then start connecting into a
network eventually reaching percolation at some lower density
threshold.

The mean intergalactic separation for our sample is $\sim 8$ Mpc
whereas we have chosen a smoothing length of 2 Mpc. The choice of a
larger smoothing length affects the measurements at the higher density
regions as discussed earlier. On the other hand, a smoothing length
smaller than the mean intergalactic separation would lead to a greater
abundance of the low-density regions that may affect some of our
results at lower density. Keeping this possibility in mind, we also
repeat our chi-square analysis considering only the results above
$\delta_{th}>3$. We find that our conclusions remain unchanged.

\section{CONCLUSIONS}

Large-scale structures like sheets and filaments are striking visual
patterns observed in galaxy distribution. The statistical significance
of these large-scale clustering patterns is difficult to judge by
visual inspections or automated identification schemes. The sheets and
filaments are anisotropic structures that introduce significant
inhomogeneity and anisotropy in the galaxy distribution. They may vary
widely in size and shape \citep{pimb04, colberg05}. However, their
large-scale distribution is expected to be uniform, provided our
Universe conforms to the cosmological principle. Any such uniformity
can only be achieved beyond the length-scale of the largest patterns
present in the galaxy distribution. In this context, the maximum size
of these large-scale structures may be treated as a lower limit for
the scale of homogeneity.

In this work, we attempt to quantify the maximum extent of the sheets
and filaments in the three-dimensional galaxy distribution for the
first time. Some analyses have been carried out for the filaments in
some earlier works \citep{bharadwaj04, pandey05, pandey11} using the
projected galaxy distributions on two-dimension. These analyses are
prone to projection effects where larger filaments can arise due to
the projection of multiple sheets and filaments.

We use the SDSS data to construct a volume-limited galaxy sample
within a cubic region and quantify the largest cluster statistic,
average filamentarity and average planarity as a function of the
density threshold. The largest cluster statistic quantifies the
connectivity in the galaxy distribution and exhibits a sudden jump at
the percolation threshold. The average filamentarity reaches a maximum
at the percolation threshold and decreases at a lower
density. The average planarity, on the other hand, increases with the
decreasing density threshold. These statistics provide a combined
picture where the galaxy distribution becomes more filamentary and
planar with decreasing density threshold. The filaments and sheets
interconnect to produce larger structures as the density threshold is
lowered. Finally, all the individual structures merge into a giant
sponge-like network spanning the entire volume.

The morphology of the large-scale structures is well described by the
Shapefinders at and above the percolation threshold. We compare the
average filamentarity and planarity in the original galaxy
distribution to those from the shuffled data and find a statistically
significant reduction in their values at all the density thresholds
above $3$. The difference between the original and the shuffled data
reduces with the increasing shuffling scale at each density
threshold. We quantify the differences using the chi-square per degree
of freedom and find that both the filaments and sheets remain
statistically significant up to the length scale of $133.5 \,
\mpc$. Our analysis is limited by the physical size of the SDSS data
cube. $133.5$ Mpc is the highest shuffling scale employed in the
present work. So the present analysis suggests that the filaments and
the sheets extend up to $\sim 130 \, \mpc$. They may even extend to
somewhat larger length scales not captured in our analysis. We do not
attempt to extrapolate our results to arrive at any definite
conclusion. The fact that the differences are more significant for
sheets than the filaments at the largest shuffling scale implies that
the sheets may extend to larger lengths than the filaments. It is
interesting to note that the result of this analysis is consistent
with the measured scale of homogeneity and isotropy from different
datasets.

We carry out our analysis in the redshift space. It is worthwhile to
mention here that the redshift space distortions (RSD) may have
significant impact on the size of the structures in our analysis. It
is well known that the structures in the real-space and redshift-space
have different power spectra, correlation functions, and pdfs
\citep{kaiser87,hamilton92,hui00}. Besides, the size of the structures
can be considerably larger in redshift space compared to the original
structures in real space
\citep{praton97,melott98}. \citet{shandarin09} show that the size of
the structures in redshift space are significantly larger in the
transverse direction than in the line of sight direction. Thus RSD can
play a substantial role in the apparent detection of structures like
`great walls'. This implies that some of the large-scale structures in
our dataset may not exist in real-space and arises due to the RSD. In
a strict sense, our results are valid in redshift space. The geometry
of the survey boundary and the shot noise can also introduce some
systematic effects on the size of the structures. In future, we plan
to use large-volume cosmological simulations to quantify the effects
of RSD, survey boundaries and shot noise on the maximum extent of the
large-scale structures.

We will be able to measure the maximum size of the sheets and
filaments in a more conclusive manner with the upcoming EUCLID survey
\citep{euclid22}. The EUCLID is expected to provide $\sim 30$ million
spectroscopic redshifts over a sky area of $15000$ square degrees and
up to a redshift of $z \sim 2$.

Finally, we conclude that the sheets and filaments in the 3D galaxy
distribution are statistically significant up to $\sim 130 \,
\mpc$. The maximum size of the filaments and sheets can be somewhat
larger but should be pretty close to this length scale. We expect to
determine these length scales using our method to deeper and larger
galaxy surveys in future.

\section{ACKNOWLEDGEMENT}
We sincerely thank an anonymous reviewer for the valuable comments and
suggestions that helped us to improve the draft. The authors thank the
SDSS team for making the data publicly available. PS acknowledges
discussions with Varun Sahni and Satadru Bag during the development of
SURFGEN2. BP would like to acknowledge financial support from the
SERB, DST, Government of India through the project CRG/2019/001110. BP
would also like to acknowledge IUCAA, Pune for providing support
through associateship programme. SS acknowledges IISER Tirupati for
support through a postdoctoral fellowship.

Funding for the SDSS and SDSS-II has been provided by the Alfred
P. Sloan Foundation, the Participating Institutions, the National
Science Foundation, the U.S. Department of Energy, the National
Aeronautics and Space Administration, the Japanese Monbukagakusho, the
Max Planck Society, and the Higher Education Funding Council for
England. The SDSS website is http://www.sdss.org/.

The SDSS is managed by the Astrophysical Research Consortium for the
Participating Institutions. The Participating Institutions are the
American Museum of Natural History, Astrophysical Institute Potsdam,
University of Basel, University of Cambridge, Case Western Reserve
University, University of Chicago, Drexel University, Fermilab, the
Institute for Advanced Study, the Japan Participation Group, Johns
Hopkins University, the Joint Institute for Nuclear Astrophysics, the
Kavli Institute for Particle Astrophysics and Cosmology, the Korean
Scientist Group, the Chinese Academy of Sciences (LAMOST), Los Alamos
National Laboratory, the Max-Planck-Institute for Astronomy (MPIA),
the Max-Planck-Institute for Astrophysics (MPA), New Mexico State
University, Ohio State University, University of Pittsburgh,
University of Portsmouth, Princeton University, the United States
Naval Observatory, and the University of Washington.

\section{DATA AVAILABILITY}
The data underlying this article are publicly available at
https://skyserver.sdss.org/casjobs/.

\bsp	
\label{lastpage}
\end{document}